\documentclass[prl,aps,preprint,amsmath,amssymb,showpacs]{revtex4}
\usepackage{graphicx}
\usepackage{epsfig}
\include{graphics}
\begin{document}

\title{Energy exchange in fast optical soliton collisions as a 
random cascade model}

\author{Avner Peleg}

\affiliation{Department of Mathematics, State University of New York
at Buffalo, Buffalo, New York 14260, USA}


\begin{abstract}
We study the dynamics of a probe soliton  
propagating in an optical fiber and exchanging energy in fast 
collisions with a random sequence of pump solitons. The  
energy exchange is induced by Raman scattering or by 
cubic nonlinear loss/gain. We show that the equation 
describing the dynamics of the probe soliton's amplitude has the 
same form as the equation for the local space average of energy 
dissipation in random cascade models in turbulence. 
We characterize the statistics of the 
probe soliton's amplitude by the $\tau_{q}$ exponents from multifractal theory 
and by the Cram\'er function $S(x)$. 
We find that the $n$th moment of the two-time 
correlation function and the bit-error-rate contribution from amplitude decay 
exhibit power-law behavior as functions of propagation distance, 
where the exponents can be expressed in terms of $\tau_{q}$ or $S(x)$.      
\end{abstract}

\pacs{05.40.-a, 42.81.Dp, 42.65.Dr, 47.53.+n} 
\maketitle

The dynamics of localized patterns in the presence of noise 
and nonlinear effects is a rich subject that is of major 
importance in many fields, including solid state 
physics \cite{Malomed89}, turbulence \cite{Frisch95}, 
and optics \cite{Agrawal2001}. Fiber optics communication 
systems, which employ optical pulses to represent information bits, 
provide an excellent example for systems where noise and nonlinearities 
have an important impact on pattern dynamics \cite{Agrawal2001}.  
It is well established that the parameters 
characterizing optical pulses in fiber optics networks can 
exhibit non-Gaussian statistics 
\cite{Menyuk95,Georges96,Falkovich2001,Ho2003,P2004,Turitsyn2005,CP2005}. 
Yet, since optical fiber systems are only 
weakly nonlinear, it was commonly believed that the statistics of 
optical pulse parameters is very different from the statistics encountered 
in strongly nonlinear systems, such as turbulence and chaotic flow, 
where intermittent dynamics exists. Two recent studies of 
pulse propagation in multichannel optical fiber transmission systems 
in the presence of delayed Raman response obtained results that stand 
in sharp contrast to this common belief \cite{P2007,CP2008}. 
Taking into account the interplay between Raman-induced 
energy exchange in pulse collisions and bit pattern 
randomness it was shown that the pulse parameters 
exhibit intermittent dynamics 
in the sense that their normalized moments grow exponentially 
with propagation distance. Furthermore, it was found that this 
intermittent dynamic behavior has important practical consequences, 
by leading to relatively large values of the bit-error-rate (BER), 
which is the probability for an error at the output of the 
fiber line.

The studies in Refs. \cite{P2007,CP2008} raise many intriguing questions, 
which are of fundamental importance in nonlinear optics, 
statistical physics, and chaos and turbulence theories.  
The main questions are: (1) Is the similarity between the fiber optics 
system and the turbulent one coincidental, or is it a consequence 
of a common underlying mechanism? (2) Is Raman scattering the only 
nonlinear process leading to intermittent dynamics of optical solitons? 
(3) Can the statistics of pulse parameters in the weakly nonlinear 
optical fiber system be analyzed by statistical mechanics 
tools that are used in the analysis of strongly nonlinear systems? 
(4) If the answer to (3) is yes, what predictions can be made 
for the main observables using these statistical mechanics tools?  
In this Letter we address these questions in detail. 
We focus attention on propagation of a probe soliton 
undergoing many collisions with a random sequence of pump 
solitons in the presence of delayed Raman response.  
First, we show that for certain setups of the pump soliton sequence, 
the dynamic equation for the probe soliton's amplitude 
has the same form as the equation for the local space average of 
energy dissipation in random cascade models in turbulence. 
The latter models play a pivotal role in the analysis of large fluctuations 
of the local energy dissipation and in obtaining 
corrections to the scaling laws of the classical Kolmogorov theory of 
turbulence \cite{Frisch95,Kolmogorov62,Novikov64,Mandelbrot74,
Benzi84,Frisch85,Halsey86,Sreenivasan87,Frisch2005}. 
Thus, our study provides a surprising answer to question 
(1): the similarity between the weakly 
nonlinear system and the strongly nonlinear one is not coincidental. 
Second, we demonstrate that Raman scattering is not the only nonlinear 
effect leading to intermittent dynamics and that 
similar behavior should be observed in 
systems where the collision-induced energy 
exchange is due to cubic nonlinear loss/gain. 
In this sense our results are quite general, since they hold for the 
three main first order perturbations to the 
nonlinear Schr\"odinger (NLS) equation that lead to 
collision-induced energy exchange.  
Third, we obtain a positive answer to question (3) by showing that 
the statistics of the probe soliton's amplitude
can be described by the $\tau_{q}$ exponents, 
which are commonly employed for analyzing multifractals and 
strange attractors in chaotic systems \cite{Benzi84,Frisch85,Halsey86}, 
and by relating the $\tau_{q}$ exponents 
to the Cram\'er function $S(x)$ via a Legendre 
transform. Fourth, we examine the implications of this dynamic behavior 
on two major observables of the probe soliton: the $n$th moment 
of the two-time equal-distance correlation function and the BER 
contribution from amplitude decay.    
We find that both observables exhibit power-law behavior as functions  
of propagation distance, where the exponents can be expressed in 
terms of $\tau_{q}$ or $S(x)$. We also reveal an intriguing similarity 
between BER dynamics and the dynamics of the $d$-measure in 
coarsening of geometrical multifractals.

It is interesting to note that several previous works studied the emergence 
of optical turbulence in nonlinear cavity ring resonators 
\cite{Ikeda80,Newell85,Akhmanov88,Akhmanov92}. 
In one of these works (Ref. \cite {Akhmanov92}) the emergence of 
turbulent-like behavior was associated with the excitation of many strongly 
interacting spatio-temporal structures, but the phenomenon was not analyzed 
quantitatively. In addition, direct mathematical links between dynamics 
in systems described by NLS and related models and dynamics 
observed in models of turbulent flow have been obtained in Refs. 
\cite{Newell91a,Newell91b,Newell2001,Zakharov2004} 
(see also references therein). However, the propagation equations in 
these studies did not take into account the effects of Raman scattering, 
and the setups considered were different from the setups considered here. 
Furthermore, the statistics of the physical observables was not analyzed 
in terms of the random cascade model and the multifractal formalism that are 
employed in the current paper.

Propagation of pulses of light in an optical fiber in the presence 
of delayed Raman response is described by the following perturbed 
NLS equation \cite{Agrawal2001}
\begin{eqnarray}
i\partial_z\Psi+\partial_t^2\Psi+2|\Psi|^2\Psi=
-\epsilon_{R}\Psi\partial_t|\Psi|^{2},
\label{cascade1}
\end{eqnarray}
where $\Psi$ is the envelope of the electric field, 
$z$ is propagation distance and $t$ is retarded time. 
The term $-\epsilon_{R}\Psi\partial_t|\Psi|^{2}$ accounts for 
the effects of delayed Raman response and $\epsilon_{R}$ 
is the Raman coefficient \cite{dimensions}.
When $\epsilon_{R}=0$, the solution of Eq. (\ref{cascade1}) 
corresponding to a single soliton with frequency $\beta$ 
is described by $\Psi_{\beta}(t,z)\!=\!
\eta_{\beta}\exp(i\chi_{\beta})\cosh^{-1}(x_{\beta})$, where
$x_{\beta}=\eta_{\beta}\left(t-y_{\beta}-2\beta z\right)$, 
$\chi_{\beta}=\alpha_{\beta}+\beta(t-y_{\beta})+
\left(\eta_{\beta}^2-\beta^{2}\right)z$, 
and $\eta_{\beta}, \alpha_{\beta}$ and $y_{\beta}$ are the 
soliton amplitude, phase and position.

Consider a single collision between a probe soliton with 
frequency $\beta=0$ and a pump soliton with frequency 
$\beta$. In a fast collision $|\beta|\gg 1$. Assuming in addition 
that $\epsilon_{R}\ll 1$, the main effect of delayed Raman 
response on the collision is an $O(\epsilon_{R})$ change 
in the soliton amplitude \cite{Chi89,Malomed91a,Kumar98,P2004,CP2005}: 
\begin{eqnarray}
\Delta\eta_{0}=
2\eta_{0}\eta_{\beta}\mbox{sgn}(\beta)\epsilon_{R}.
\label{cascade2}
\end{eqnarray}   
The effect of the collision in order $\epsilon_{R}/\beta$ 
is a frequency shift given by \cite{Chi89,Kumar98,P2004,CP2005}:
$\Delta\beta_{0}=-(8\eta_{0}^{2}\eta_{\beta}\epsilon_{R})/(3|\beta|)$. 
This effect and effects of order $\epsilon_{R}^{2}$ and higher   
can be neglected for the dynamical setups considered below. 
In addition, for other types of perturbations, such as those due 
to third order dispersion, the collision-induced  
changes in amplitude and frequency are of higher order in both the 
parameter $\epsilon$ characterizing the perturbative process 
and $1/|\beta|$ (see, e.g., Refs. \cite{Malomed91b,PCG2003,PCG2004}).

We now describe propagation of a probe soliton under many 
collisions with a random sequence of pump solitons. The pump 
solitons are located at time slot centers, and each time slot 
can be either occupied or empty. The occupation state of the 
$j$th time slot is described by the random variable $\zeta_{j}$: 
$\zeta_{j}=1$ with probability $s$ if the slot is occupied 
and $\zeta_{j}=0$ with probability $1-s$ otherwise. 
It is assumed that different time slots are uncorrelated: 
$\langle\zeta_{i}\zeta_{j}\rangle=s^{2}$ if $i\ne j$.
The probe soliton is initially located at $y_{0}(0)=0$ and  
the frequencies of the pump solitons are $\beta_{j}=\Delta\beta>0$.  
Since we look for power-law behavior of the physical 
observables we assume that the pump solitons are initially located at 
$y_{\beta j}(0)=-a^{j}T$, where $T$ and $a>1$ are constants.
Therefore, the collision between the $j$th pump soliton 
and the probe soliton occurs at a distance $z_{j}$,  
given by: $z_{j}=a^{j}\Delta z_{c}^{(1)}$, 
where $\Delta z_{c}^{(1)}=T/(2\Delta\beta)$.

\begin{figure}[ptb]
\epsfxsize=7.0cm  \epsffile{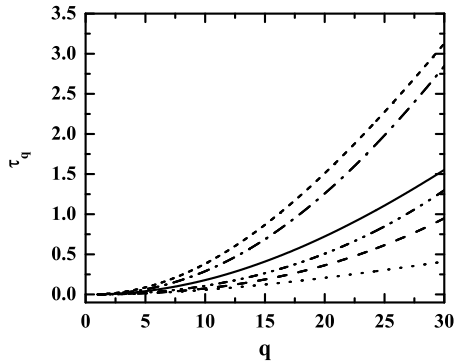}
\caption{The $\tau_{q}$ curve for the probe soliton's amplitude 
for $\epsilon_{R}=0.03$ (pulse width = 0.2 ps), $a=1.25$, 
and $\eta_{\beta}(0)=1$. 
The solid, dashed and dotted lines represent 
the values obtained in the case where $\eta_{\beta j}(0)$ is  
deterministic with $s=0.5$, $s=0.1$, and $s=0.9$, respectively. 
The dashed-dotted, dashed-dotted-dotted, and short-dashed curves 
correspond to $\tau_{q}$ in the case where $\eta_{\beta j}(0)$ is random 
with $s=0.5$, $\Delta\eta_{\beta j}=0.5$, and $\rho=0.5$, 
$\rho=0.1$, $\rho=0.9$, respectively.}
\label{fig1}
\end{figure}

Using Eq. (\ref{cascade2}) and assuming that 
$\eta_{\beta j}(0)=\eta_{\beta}(0)$ for all pump solitons 
we find that the probe soliton's 
amplitude after $J$ collisions is 
$\eta_{0}(z_{J})=\eta_{0}(z_{J-1})[1+2\epsilon_{R}\eta_{\beta}(0)\zeta_{J}]$, 
leading to $\eta_{0}(z_{J})=\eta_{0}(0)\prod_{j=1}^{J}W_{j}$, 
where $W_{j}=1+2\epsilon_{R}\eta_{\beta}(0)\zeta_{j}$. Compensation of  
average cross talk effects can be introduced in a 
straightforward manner. In this case the  probe soliton's amplitude 
after $J$ collisions is
$\eta_{0}(z_{J})=\eta_{0}(z_{J-1})
[1+2\epsilon_{R}\eta_{\beta}(0)(\zeta_{J}-s)]$, 
resulting in 
\begin{eqnarray}
\eta_{0}(z_{J})=\eta_{0}(0)\prod_{j=1}^{J}\tilde W_{j},
\label{cascade3}
\end{eqnarray} 
where $\tilde W_{j}=1+2\epsilon_{R}\eta_{\beta}(0)(\zeta_{j}-s)$. 
Notice that $\langle\tilde W_{j}\rangle=1$, and therefore 
$\langle\eta_{0}(z_{J})\rangle=1$.  
Equation (\ref{cascade3}) has the 
same form as the equation for the local space average of 
energy dissipation in random cascade models in turbulence. 
(Compare with Ref. \cite{Frisch95}, p. 166). In this equivalence, 
$z_{J}$, $\Delta z_{c}^{(1)}$ and $\eta_{0}(z_{J})$, 
play the roles of eddy size $l$, 
upper turbulence cutoff $l_{0}$, and energy 
dissipation of eddies of size $l$, $\varepsilon_{l}$, 
respectively. The evolution of both $\eta_{0}$ and 
$\varepsilon_{l}$ is multiplicative and dissipative. 
Notice that in the fiber optics system energy cascades between 
successive collisions, whereas in the turbulent model 
the cascade is from large eddies to smaller ones.    
Using Eq. (\ref{cascade3}) and the statistical independence 
of the $\tilde W_{j}$ factors, we obtain 
\begin{eqnarray}
\langle \eta_{0}^{q}(z_{J})\rangle=\eta_{0}^{q}(0)
\left(z_{J}/\Delta z_{c}^{(1)}\right)^
{\log_{a}\langle \tilde W^{q}\rangle}.
\label{cascade5}
\end{eqnarray}
We define the $\tau_{q}$ moments in a similar manner to 
Refs. \cite{Benzi84,Frisch85,Halsey86}: 
$\tau_{q}\equiv\log_{a}\langle \tilde W^{q}\rangle$.    
For the fiber optics system: 
\begin{eqnarray}
\langle \tilde W^{q}\rangle= 
sw_{1}^{q}+(1-s)w_{2}^{q},
\label{cascade7}
\end{eqnarray}         
where $w_{1}=1+2(1-s)\epsilon_{R}\eta_{\beta}(0)$ and 
$w_{2}=1-2s\epsilon_{R}\eta_{\beta}(0)$. 
The $\tau_{q}$ curve obtained by using Eq. (\ref{cascade7}) is 
plotted in Fig. \ref{fig1} for $\epsilon_{R}=0.03$ 
($\tilde\tau_{0}=0.2$ ps), $a=1.25$, $\eta_{\beta}(0)=1$, 
and different $s$-values.  

We emphasize that similar dynamics of the probe soliton's amplitude 
is expected in systems described by perturbed 
NLS equations, where the perturbation is due to cubic nonlinear 
loss/gain. For these systems the 
$-\epsilon_{R}\Psi\partial_t|\Psi|^{2}$ term  
on the right hand side of Eq. (\ref{cascade1}) is replaced 
by $\mp\epsilon_{c}|\Psi|^{2}\Psi$, respectively, where 
$\epsilon_{c}$ is the cubic nonlinear loss/gain.     
The main effect of a fast collision in this case is an  
amplitude change, which is given by an equation of the form 
(\ref{cascade2}) with $\mbox{sgn}(\beta)\epsilon_{R}$ replaced 
by $\mp 2\epsilon_{c}/|\beta|$. Thus, the results of this Letter 
are quite general since they hold for Raman scattering and 
nonlinear loss/gain, which are the three main first order perturbations 
to the NLS equation that lead to energy exchange in pulse 
collisions.

It is possible to relate the $\tau_{q}$ exponents 
to the Cram\'er function $S(x)$, 
characterizing the statistics of the pump soliton  
bit pattern, by a method similar to the one used 
for random cascade models and multifractal sets 
\cite{Benzi84,Frisch85,Halsey86}. For this purpose we notice 
that the variable $m=\sum_{j=1}^{J}\zeta_{j}$ is binomially 
distributed with probability density function 
(PDF) $P(m;J)=[J!s^{m}(1-s)^{J-m}]/[m!(J-m)!]$. 
Using Stirling's formula we obtain     
\begin{eqnarray}
P(m;J)\simeq 
\left[-S''(x)\right]^{1/2}\exp[JS(x)]/
\left(2\pi J\right)^{1/2},
\label{cascade8}
\end{eqnarray}   
where $x=m/J$ and 
\begin{eqnarray}
S(x)=-x\ln(x/s)-(1-x)\ln[(1-x)/(1-s)].
\label{cascade10}
\end{eqnarray}   
To obtain the relation between $\tau_{q}$ 
and $S(x)$ we notice that 
$a^{J\tau_{q}}=\langle\prod_{j=1}^{J}\tilde W_{j}^{q}\rangle$,   
and consequently: 
$a^{J\tau_{q}}=\langle\exp\left\{q\sum_{j=1}^{J}
\ln\left[1+2\epsilon_{R}(\zeta_{j}-s)\right]\right\}\rangle$.
Expressing the sum in the exponent's argument in terms of 
$x$, $w_{1}$ and $w_{2}$ and using the large-deviations 
theorem we obtain 
\begin{eqnarray} &&
\!\!\!\!\!\! a^{J\tau_{q}}\simeq
\int \mbox{d}x \left[\frac{-JS''(x)}{2\pi}\right]^{1/2}
\exp\left\{qJ\left[x\ln(w_{1}/w_{2})
\right.
\right.
\nonumber \\&&
\left.
\left.
+\ln(w_{2})\right]+JS(x)\right\}.
\label{cascade13}
\end{eqnarray}     
Employing a saddle point approximation we obtain 
the Legendre transform relating $\tau_{q}$ and $S(x)$: 
\begin{eqnarray}
\tau_{q}=
\left\{q\left[x\ln(w_{1}/w_{2})+\ln(w_{2})\right]+S(x)\right\}/\ln a,
\label{cascade14}
\end{eqnarray}    
where $x(q)$ is determined by the conditions
$q\ln(w_{1}/w_{2})=-S'(x)$, and $S''(x)<0$.

Since $\epsilon_{R}\ll 1$ the $\tilde W_{j}$ 
factors in the fiber optics system are only slightly different 
from 1. It is therefore interesting to look for ways to enhance 
the randomness of the collision-induced energy exchange. 
Here we describe one disorder enhancement 
mechanism, which also leads to a natural 
generalization of the model described by Eq.  
(\ref{cascade3}). Consider a setup in which 
the amplitude of the $j$th pump soliton is given by
$\eta_{\beta j}(0)=\eta_{\beta}(0)\pm\Delta\eta_{\beta}$ with probabilities  
$\rho$ and $1-\rho$, respectively. A straightforward analysis shows that 
the $\eta_{0}$ dynamics is still described by 
Eq. (\ref{cascade3}). However, the $\tilde W_{j}$ factors 
can now attain three values: 
$w_{1}=1+2\epsilon_{R}[(1-s)\eta_{\beta}(0)+(1-2\rho)s\Delta\eta_{\beta}+
\Delta\eta_{\beta}]$ with probability $\rho s$, 
$w_{2}=1+2\epsilon_{R}[(1-s)\eta_{\beta}(0)+(1-2\rho)s\Delta\eta_{\beta}-
\Delta\eta_{\beta}]$ with probability $(1-\rho)s$, and 
$w_{3}=1+2\epsilon_{R}[-s\eta_{\beta}(0)+(1-2\rho)s\Delta\eta_{\beta}]$ 
with probability $1-s$. The average of $\tilde W^{q}$ is 
\begin{eqnarray}
\langle \tilde W^{q}\rangle= 
\rho sw_{1}^{q}+(1-\rho)sw_{2}^{q}+
(1-s)w_{3}^{q}. 
\label{cascade16}
\end{eqnarray}        
The $\tau_{q}$ curve for this model is shown in Fig. \ref{fig1}. 
One can see that $\tau_{q}$ attains larger values when 
$\eta_{\beta j}(0)$ is random for $s=0.5$, $\rho=0.5$, and $\rho=0.9$, 
compared with the case where $\eta_{\beta j}(0)$ is deterministic 
and $s=0.5$. Thus, the cumulative effect of energy exchange 
in the collisions is indeed enhanced by randomness of 
$\eta_{\beta j}(0)$. To obtain the Legendre transform 
between $\tau_{q}$ and $S(x)$, 
we denote by $m_{i}, i=1,2,3$, the number of occurrences 
of pump solitons with $\tilde W_{j}=w_{i}$. 
The PDF of $m_{1}$ and $m_{2}$ is trinomial: 
$P(m_{1},m_{2};J)=\{J!(\rho s)^{m_{1}}[(1-\rho)s]^{m_{2}}
(1-s)^{J-m_{1}-m_{2}}]/[m_{1}!m_{2}!(J-m_{1}-m_{2})!]$. 
Using Stirling's formula we arrive at
\begin{eqnarray}
P(m_{1},m_{2};J)\simeq
\frac{J^{1/2}\exp[J\tilde S(x_{1},x_{2})]}
{2\pi\left[m_{1}m_{2}(J-m_{1}-m_{2})\right]^{1/2}},
\label{cascade17}
\end{eqnarray}   
where $x_{1,2}=m_{1,2}/J$, and  
\begin{eqnarray}&&
\tilde S(x_{1},x_{2})=-x_{1}\ln[x_{1}/(\rho s)]-
x_{2}\ln\{x_{2}/[(1-\rho)s]\}
\nonumber \\&&
-(1-x_{1}-x_{2})\ln[(1-x_{1}-x_{2})/(1-s)].
\label{cascade19}
\end{eqnarray}   
By employing the large-deviations 
theorem and a saddle point calculation we obtain the following 
generalization of the Legendre transform given by 
Eq. (\ref{cascade14}):  
\begin{eqnarray} &&
\tau_{q}=
\left\{q\left[x_{1}\ln(w_{1}/w_{3})+
x_{2}\ln(w_{2}/w_{3})+
\ln(w_{3})\right]
\right.
\nonumber \\&&
\left.
+\tilde S(x_{1},x_{2})\right\}/\ln a,
\label{cascade21}
\end{eqnarray}    
where $x_{1}(q)$ and $x_{2}(q)$ are determined by 
\begin{eqnarray}
q\ln(w_{i}/w_{3})=-\frac{\partial\tilde S}
{\partial x_{i}}\left.\right|_{x_{i}=x_{i}(q)}  
,\;\;\; i=1,2,
\label{cascade22}
\end{eqnarray}   
$\partial^{2}\tilde S/\partial^{2} x_{1}<0$, and 
$(\partial^{2}\tilde S/\partial^{2} x_{1})
(\partial^{2}\tilde S/\partial^{2} x_{2})-
(\partial^{2}\tilde S/\partial x_{1} \partial x_{2})^{2}>0$.  
Since the exact expression for $\langle\tilde W^{q}\rangle$
[Eq. (\ref{cascade16})] can be retrieved by using 
Eqs. (\ref{cascade19})-(\ref{cascade22}) we conclude 
that the $\tau_{q}$ exponents give a correct characterization 
of the statistics of $\eta_{0}$. In addition, 
Eqs. (\ref{cascade19})-(\ref{cascade22})
can be further generalized for the case where the $\tilde W_{j}$
factors attain any finite number of values.

We now discuss implications of the dynamics of the probe 
soliton's amplitude. We start by considering the $n$th moment 
of the two-time equal-distance correlation:
\begin{eqnarray} &&
C_{01}^{(n)}=\langle\left(\eta_{00}\eta_{01}
\right)^{n}\rangle/
(\langle\eta_{00}\rangle^{n}
\langle\eta_{01}\rangle^{n}).
\label{cascade25}
\end{eqnarray}   
The function $C_{01}^{(n)}$ measures correlation between the 
amplitudes $\eta_{00}$ and $\eta_{01}$ of two probe solitons, 
whose initial positions are $y_{00}(0)=0$ and $y_{01}(0)=AT$, 
respectively, where $A$ is a constant. We remark that high moments  
of velocity correlation function along a given direction play 
an important role in turbulence theory \cite{Frisch95}. 
From Eq. (\ref{cascade3}) it follows that 
$\langle\left[\eta_{00}(z_{J})\eta_{01}(z_{J})\right]^{n}\rangle=
\eta_{00}^{n}(0)\eta_{01}^{n}(0)
\langle\prod_{j=1}^{J}\tilde W_{j0}^{n}
\prod_{j=1}^{j_{max}}\tilde W_{j1}^{n}\rangle$,
where $j_{max}$ is the number of collisions experienced 
by the 01 probe soliton, and $\tilde W_{j1}=\tilde W_{j0}$ 
for $j=1,\dots, j_{max}$. Using this relation, the 
statistical independence of the $\tilde W_{j0}$ factors 
and the definition of the $\tau_{q}$ exponents,  
we obtain
\begin{eqnarray} &&
C_{01}^{(n)}(z_{J})\simeq\left[\frac{z_{J}-z_{11}}
{\Delta z_{c}^{(1)}}\right]^{\tau_{2n}}
\left[\frac{z_{J}}{z_{J}-z_{11}}\right]^{\tau_{n}}, 
\label{cascade28}
\end{eqnarray}  
where $z_{11}=A\Delta z_{c}^{(1)}$. Thus, the $n$th 
moment of the two-time correlation function is 
a product of power-laws of two different scaled distances.

\begin{figure}[ptb]
\epsfxsize=7.0cm  \epsffile{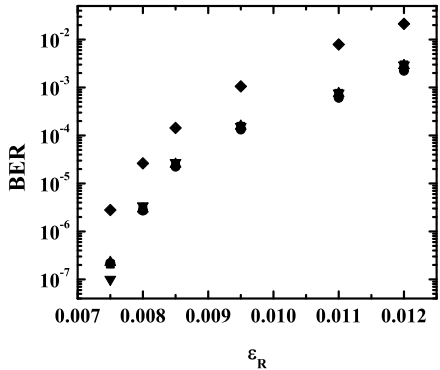}
\caption{Total BER and BER contribution due to pulse decay, 
$\mbox{BER}_{\eta}$, vs the Raman coefficient $\epsilon_{R}$ 
for $\eta_{0}(0)=1$, $\eta_{\beta}(0)=2$, $a=1.25$, $T=5$, and 
$\beta_{2}=-4\mbox{ps}^{2}/\mbox{km}$.
The squares, up triangles, and circles stand for the exact result 
for $\mbox{BER}_{\eta}$, the approximate expression given 
by Eq. (\ref{cascade30}), and the first term on the right 
hand side of Eq. (\ref{cascade30}), respectively.
The down triangles and diamonds represent the total BER for 
$\Delta\beta=40$ and $\Delta\beta=20$, respectively.}
\label{fig2}
\end{figure}

One of the most important quantities characterizing the 
performance of fiber optics systems is the BER. Since the BER is often 
determined by the tail of the PDF of the pulse parameters 
one can expect that it would be closely related to the   
Cram\'er function. We show that this is indeed the case 
for the system described here. We consider the setup where 
the amplitudes of the pump solitons are deterministic 
[$\eta_{\beta j}(0)=\eta_{\beta}(0)$] and focus attention on 
the contribution to the probe soliton's BER due to 
amplitude decay, $\mbox{BER}_{\eta}$. 
This contribution is defined by: 
$\mbox{BER}_{\eta}\equiv\int_{0}^{\eta_{th}}
\mbox{d}\eta_{0}F(\eta_{0})$, where $F(\eta_{0})$ is the 
amplitude PDF and $\eta_{th}$ is the threshold for an error. 
Using $\eta_{0}(z_{J})=\eta_{0}(0)w_{1}^{m}w_{2}^{J-m}$ 
and Eq. (\ref{cascade8}) we obtain  
\begin{eqnarray} &&
\mbox{BER}_{\eta}\simeq
\sum_{m=0}^{m_{th}}\frac{J^{1/2}\exp[JS(m/J)]}
{\left[2\pi m(J-m)\right]^{1/2}},
\label{cascade29}
\end{eqnarray}             
where $m_{th}$ is the solution of $\eta_{0}(m)=\eta_{th}$. 
Since $S(x)$ is an increasing function of $x$  
the main contribution to the sum on the right hand side 
of Eq. (\ref{cascade29}) comes from a close neighborhood 
of $m_{th}$. Taking into account the two leading 
terms we arrive at: 
\begin{eqnarray}&&
\!\!\!\!\!\!\!\!\!\!\!\!\!\!\!\!
\mbox{BER}_{\eta}\simeq
\!\left[\frac{-S''(x)}{2\pi J}\right]^{1/2}
\!\left[\frac{z_{J}}{\Delta z_{c}^{(1)}}\right]
^{\frac{S(x_{th})}{\ln a}}
\!\left[1+a^{-\frac{S'(x_{th})}{\ln a}}\right]\!,
\label{cascade30}
\end{eqnarray}    
where $x_{th}=m_{th}/J$. Thus, in the leading order 
$\mbox{BER}_{\eta}$ grows like a power law with propagation distance, 
where the exponent is the value of the Cram\'er function 
at the error threshold. The $\epsilon_{R}$-dependence of  
$\mbox{BER}_{\eta}$ in the range $0.0075\le \epsilon_{R}\le 0.012$ 
($0.5\mbox{ps}\le\tilde\tau_{0}\le 0.8\mbox{ps}$) is shown in 
Fig. \ref{fig2} for $\eta_{0}(0)=1$, $\eta_{\beta}(0)=2$, 
$\eta_{th}=0.7$ and $J=27$. 
For the choice $a=1.25$, $T=5$, $\Delta\beta=40$, 
and $\beta_{2}=-4\mbox{ps}^{2}/\mbox{km}$, 
for example, we obtain  $7.96\mbox{THz}<\Delta\nu<12.74\mbox{THz}$ 
for the frequency difference, and $1.07\mbox{km}<X_{27}<2.75\mbox{km}$ 
for the propagation distance. 
It is seen that both Eq. (\ref{cascade30}) and the first term on 
the right hand side of Eq. (\ref{cascade30}) are good 
approximations to the exact result over a wide range 
of $\mbox{BER}_{\eta}$ values. It is interesting that the 
dynamic behavior of $\mbox{BER}_{\eta}$ is very similar to the 
behavior of the $d$-measure of geometrical 
multifractals during the late stage of coarsening
(compare Eq. (\ref{cascade29}) with Eq. (9) in Ref. \cite{PM2000}).

In many cases the dynamics of the other three soliton 
parameters is coupled to the amplitude dynamics and  
as a result, these parameters can be strongly influenced by 
amplitude fluctuations. For the system considered here the 
most important effect is due to the Raman-induced self 
frequency shift, which is given by \cite{Gordon86}:
$\beta_{0}(z)=
-(8\epsilon_{R}/15)\int_{0}^{z}\mbox{d}z'
\eta_{0}^{4}(z')$. 
This frequency shift leads to a position shift:
\begin{eqnarray} &&
y_{0}(z)= 
-\frac{16\epsilon_{R}}{15}
\int_{0}^{z}\mbox{d}z'\int_{0}^{z'}\mbox{d}z''\eta_{0}^{4}(z'')
\label{cascade31}
\end{eqnarray}
that can give significant contribution to the probe soliton's 
total BER. It is therefore important to evaluate the impact of 
this process on the total BER in comparison with the contribution 
$\mbox{BER}_{\eta}$ coming solely from amplitude decay. 
Since we do not have an analytic expression for the PDF of $y$ we 
carry out Monte Carlo simulations with Eqs. (\ref{cascade3}) 
and (\ref{cascade31}). We define the relative position shift 
$\tilde y_{0}=y_{0}-\langle y_{0}\rangle$, where the average 
$\langle y_{0}\rangle$ is assumed to be compensated by filters. 
For each realization of the $W_{j}$ factors we compute the total 
energy at the detector at distance $z_{J}$: 
\begin{eqnarray} &&
\!\!\!\!\!\!\!\!
I(z_{J})=\eta_{0}(z_{J})\left\{
\tanh[\eta_{0}(z_{J})(T/2-\tilde y_{0}(z_{J}))]+
\tanh[\eta_{0}(z_{J})(T/2+\tilde y_{0}(z_{J})]\right\}.
\label{cascade32}
\end{eqnarray}       
An occupied time slot is considered to be in error, if 
$I(z_{J})<I_{th}=2\eta_{th}\tanh(\eta_{th}T/2)$. 
We use the same parameter values as described in the previous paragraph, 
but with two different values of $\Delta\beta$: 
$\Delta\beta=40$ and $\Delta\beta=20$ \cite{BER_eta}. 
The total BER obtained in the simulations is shown in Fig. \ref{fig2}.  
It is seen that for $\Delta\beta=40$ the total BER is very close to 
$\mbox{BER}_{\eta}$, that is, error generation is dominated by 
amplitude decay. In contrast, for $\Delta\beta=20$, 
error generation is dominated by the Raman-induced position shift, 
and as a result the total BER is much larger than $\mbox{BER}_{\eta}$.  
These results can be explained by noting that the smaller inter-collision 
distances for $\Delta\beta=40$ lead to a smaller total propagation distance, 
and consequently, the position shift is relatively small. 
For $\Delta\beta=20$, the inter-collision distances and the total propagation 
distance are large, resulting in relatively large position shifts.

In summary, we studied the amplitude dynamics of a probe NLS soliton, 
exchanging energy in fast collisions with a random sequence of pump 
solitons. We showed that the equation for the probe soliton's 
amplitude has the same form as the equation for the local space average 
of energy dissipation in random cascade models in turbulence.  
We found that the $n$th moment of the two-time correlation function and 
the BER contribution from amplitude decay exhibit power-law 
behavior as functions of propagation 
distance, where the exponents can be expressed in terms 
of the $\tau_{q}$ exponents or the Cram\'er function.
Thus, our study provides a surprising and very useful perspective 
on the relation between disorder effects on weakly nonlinear 
systems described by perturbed NLS equations, and strongly nonlinear 
systems, such as turbulent flow.


\begin{thebibliography}{}
\bibitem{dimensions} The dimensionless $z$ in Eq. (\ref{cascade1}) is
$z=(|\beta_{2}| X)/(2\tilde\tau_{0}^{2})$, where $X$ is the actual position,
$\tilde\tau_{0}$ is the soliton width, and $\beta_{2}$ is the second order
dispersion coefficient. The dimensionless retarded time is
$t=\tilde\tau/\tilde\tau_{0}$, where $\tilde\tau$ is the retarded time.  
The spectral width is $\nu_{0}=1/(\pi^{2}\tilde\tau_{0})$ and the 
frequency difference is $\Delta\nu=(\pi\Delta\beta\nu_{0})/2$. 
The coefficient $\epsilon_{R}$ is given by 
$\epsilon_{R}=0.006/\tilde\tau_{0}$, where $\tilde\tau_{0}$ is 
in picoseconds.
\bibitem{BER_eta} Notice that once $J$ is specified the value of 
$\mbox{BER}_{\eta}$ is independent of $\Delta\beta$. Therefore, 
the $\mbox{BER}_{\eta}$ values are the same for $\Delta\beta=20$ 
and $\Delta\beta=40$.
\end{thebibliography}
\end{document}